\documentclass[aps,prb,twocolumn,amsmath,amssymb,superscriptaddress,reprint,longbibliography]{revtex4-2}

\usepackage[utf8]{inputenc}
\usepackage{graphicx}
\usepackage{color}
\usepackage[colorlinks=true,citecolor=blue]{hyperref}
\usepackage{units}
\usepackage{orcidlink}

\newcommand{\bra}[1]{\langle #1|}
\newcommand{\ket}[1]{|#1\rangle}

\newcommand{\up}{\uparrow}
\newcommand{\down}{\downarrow}
\renewcommand{\vec}[1]{\mathbf{#1}}

\newcommand{\kBT}{k_\text{B}T}
\DeclareMathOperator{\re}{Re}
\DeclareMathOperator{\im}{Im}

\begin{document}
\title{Phase-dependent transport in thermally-driven superconducting single-electron transistors}
\author{Alexander G. Bauer}
\affiliation{Theoretische Physik, Universität Duisburg-Essen and CENIDE, D-47048 Duisburg, Germany}
\author{Björn Sothmann\,\orcidlink{0000-0001-9696-9446}}
\affiliation{Theoretische Physik, Universität Duisburg-Essen and CENIDE, D-47048 Duisburg, Germany}
\date{\today}

\begin{abstract}
We investigate thermally-driven transport of heat and charge in a superconducting single-electron transistor by means of a real-time diagrammatic transport theory. Our theoretical approach allows us to account for strong Coulomb interactions and arbitrary nonequilibrium conditions while performing a systematic expansion in the tunnel coupling. We find that a temperature bias across the system gives rise to finite heat and charge currents close to the particle-hole symmetric point which depend both on the gate voltage as well as on the phase difference between the superconducting reservoirs. The finite thermoelectric effect arises due to level renormalization from virtual tunneling processes.
Furthermore, we find that the phase bias can give rise to finite charge currents even in the presence of an inversion-symmetric temperature bias.
\end{abstract}

\maketitle
\section{\label{sec:intro}Introduction}
The ongoing miniaturization of electrical circuits leads to ever more powerful computers. At the same time, it also gives rise to an increased production of waste heat that has to be managed, e.g., by active cooling. Therefore, understanding in details the principles that govern the transport of electrical charge and heat at the nanoscale are of crucial importance for future technological advances~\cite{giazotto_opportunities_2006}.

During the last decade, the investigation of phase-coherent heat transport in superconducting nanostructures has received a lot of attention. It dates back to the theoretical prediction that the thermal conductance of a Josephson junction depends on the superconducting phase difference due to Andreev-like processes in which quasiparticles above the superconducting gap are converted from electron-like to hole-like and vice versa while creating or annihilating Cooper pairs~\cite{maki_entropy_1965,maki_entropy_1966,guttman_phase-dependent_1997,zhao_phase_2003,zhao_heat_2004}. This phase-dependent contribution to the thermal conductance has been observed experimentally only recently~\cite{giazotto_josephson_2012} because measuring heat currents at the nanoscale is challenging. Since then, the field of phase-coherent caloritronics has seen a tremendous progress, both experimentally with the observation of heat diffraction~\cite{martinez-perez_quantum_2014}, and the realization of thermal diodes~\cite{martinez-perez_rectification_2015}, heat modulators~\cite{fornieri_nanoscale_2016}, $0-\pi$ thermal Josephson junctions~\cite{fornieri_0-pi_2017}, and thermal routers~\cite{timossi_phase-tunable_2018} but also with theoretical proposals for thermometers~\cite{giazotto_ferromagnetic-insulator-based_2015,guarcello_nonlinear_2019}, thermal transistors~\cite{fornieri_negative_2016}, refrigerators~\cite{solinas_microwave_2016,marchegiani_-chip_2018}, superconducting heat engines~\cite{marchegiani_self-oscillating_2016,vischi_coherent_2017}, temperature amplifiers~\cite{paolucci_phase-tunable_2017}, thermal memory~\cite{guarcello_josephson_2018}, photon detectors~\cite{solinas_proximity_2018,heikkila_thermoelectric_2018,virtanen_josephson_2018}, thermal switches~\cite{sothmann_high-efficiency_2017}, rectifiers~\cite{giazotto_thermal_2013,bours_phase-tunable_2019}, and heat circulators~\cite{hwang_phase-coherent_2018}, see also Refs.~\cite{fornieri_towards_2017,hwang_phase-coherent_2020} for recent reviews.
Apart from electronic heat transport, there has also been a number of works devoted to photonic heat transport in superconducting circuits in the context of circuit quantum electrodynamics both from a theoretical~\cite{ojanen_mesoscopic_2008,karimi_otto_2016,thomas_photonic_2019,xu_heat_2021} as well as from an experimental~\cite{ronzani_tunable_2018,senior_heat_2020,maillet_electric_2020} perspective.

So far, the study of phase-coherent caloritronics has mainly been restricted to systems in which interaction effects can be treated on a mean-field level or neglected altogether.
The influence of strong Coulomb interactions on phase-dependent thermal transport through a single-level quantum dot weakly tunnel coupled to two superconducting reservoirs has been investigated in Ref.~\cite{kamp_phase-dependent_2019}. 
Here, we consider thermally driven transport through a superconducting single-electron transistor consisting of a small superconducting island tunnel coupled to two superconducting reservoirs. Previously, voltage-driven charge transport has been investigated theoretically in superconducting single-electron transistors with a focus on average currents~\cite{van_den_brink_combined_1991,matveev_parity-induced_1993,siewert_charge_1996,leppakangas_effect_2008,cole_parity_2015} and current fluctuations~\cite{kack_full-frequency_2003,kirton_charge_2010}. Superconducting single-electron transport have also been realized experimentally~\cite{joyez_observation_1994,amar_2e_1994,eiles_combined_1994,matters_influence_1995,hadley_3e_1998,corlevi_coulomb_2006,van_woerkom_one_2015} with a special emphasis on parity effects which can be important for future superconducting qubit technology.

Compared to a single-level quantum dot coupled to superconducting reservoirs, a superconducting single-electron transistor offers a number of additional features. Since the island is a superconductor itself, it provides an additional phase degree of freedom. Furthermore, it has a BCS density of states which depends on energy in a nontrivial fashion. As the island contains a macroscopic number of electrons, it has a well-defined temperature which can in general be different from the temperatures of the superconducting reservoirs. The complex interplay between strong Coulomb interactions on the island, Andreev processes at the tunnel junctions and a nonequilibrium scenario induced by a temperature bias gives rise to a number of interesting transport phenomena that occur already to first order in the island-reservoir coupling. In particular, we find that both, the thermally driven charge and heat current depend in a nontrivial manner on the gate voltage as well as on the phase difference across the single-electron transistor. Remarkably, a finite thermoelectric effect occurs close to the particle-hole symmetric point due to level renormalizations on the island induced by virtual tunneling processes. Finally, we unveil that finite charge currents can arise in the presence of an inversion-symmetric temperature bias when a finite phase bias is applied across the system.

The paper is organized as follows. We describe our model in Sec.~\ref{sec:model}. In Sec.~\ref{sec:method}, we introduce a real-time diagrammatic approach which we use to calculate the transport properties of our system. Our results are presented in Sec.~\ref{sec:results}. Conclusions are drawn in Sec.~\ref{sec:conclusions}.

\section{\label{sec:model}Model}
\begin{figure}
		\includegraphics[width=1.0\columnwidth]{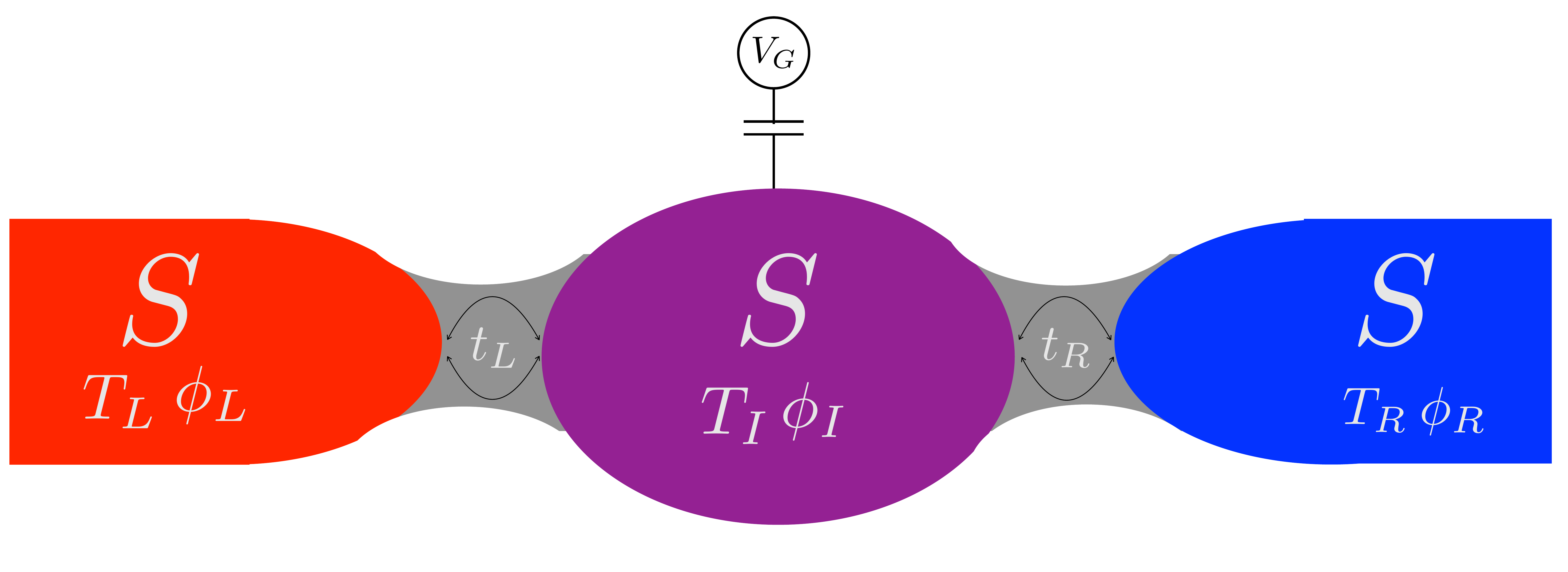}
	\caption{\label{fig:Setup}Setup of a superconducting island  weakly coupled to two superconducting leads. In general, all three parts of the junction are allowed to have different temperatures and superconducting phases. Additionally, the equilibrium occupation of the island can be controlled via the gate voltage  $V_G$ of the island. }
\end{figure}
We consider a small superconducting island weakly tunnel coupled to two superconducting reservoirs, see Fig.~~\ref{fig:Setup}. All three superconducting elements are kept at the same electrochemical potential but are allowed to have different temperatures $T_\eta$, $\eta=\text{L,R,I}$. This gives rise to a nonequilibrium situation in which heat and charge currents can flow through the system. The system is described by a total Hamiltonian of the form $H=\sum_{\eta}H_\eta+H_\text{ch}+H_\text{tun}$. Here, the first term describes the two superconducting reservoirs, $\eta=\text{L,R}$ as well as the island $\eta=\text{I}$ in terms of a conventional BCS mean-field Hamiltonian of the form
\begin{equation}
	H_\eta=\sum_{\vec k\sigma}\varepsilon_{\eta\vec k}a_{\eta\vec k\sigma}^\dagger a_{\eta\vec k\sigma}+\Delta_\eta e^{i\phi_\eta}\sum_{\vec k}a_{\eta -\vec k\up}a_{\eta\vec k\down}+\text{H.c.}
\end{equation}
While the first part describes the kinetic energy of electrons, the second part describes the superconducting pairing of electrons and contains the absolute value $\Delta_\eta$ of the superconducting order parameter and its phase $\phi_\eta$. The temperature dependence of the order parameter follows from a self-consistency equation and can be approximated for all temperatures below the critical temperature $T_c$ with high accuracy by
\begin{equation}
	\Delta_\eta(T)=\Delta_{0}\tanh\left(1.74\sqrt{\frac{T_c}{T_\eta}-1}\right),
\end{equation}
where $\Delta_{0}$ is the absolute value of the order parameter at zero temperature which is linked to the critical temperature by $\kBT_{c}=0.568\Delta_{0}$  where we assumed that all three superconductors are  made from the same material and therefore have the same critical temperature.

The superconductivity-related part of the Hamiltonian can be diagonalized by a Bogoliubov transformation. The resulting eigenstates are given by linear combinations of electrons and holes and have a dispersion relation given by $E_{\eta\vec k}=\sqrt{\varepsilon_{\eta\vec k}^2+\Delta_\eta^2}$. The associated BCS density of states normalized to the reservoir density of states in the normal state $\rho_{\text{N},\eta}$ is given by 
\begin{equation}
	\rho_\eta(\omega)=\left|\re\left[\frac{\omega+i\Gamma}{\sqrt{(\omega+i\Gamma)^2-\Delta_\eta^2}}\right]\right|
	\label{eq:rhoDynes}
\end{equation}
where  $\Gamma$  denotes the Dynes parameter which accounts phenomenologically for a finite life time of quasiparticle excitations in the superconductors.

The second part of the total Hamiltonian,
\begin{equation}
	H_\text{ch}=E_C\left(\sum_{\mathbf{k} \sigma}a_{\text{I}\vec k\sigma}^\dagger a_{\text{I}\vec k\sigma}-n_g\right)^2,
\end{equation}
describes charging effects on the island. Here, $E_C=e^2/(2C)$ is the charging energy that is required to put excess electrons on the island. Since the total capacitance $C$ of the island is small, the charging energy is large and impacts transport via Coulomb-blockade phenomena. The equilibrium occupation of the island $n_g$ can be tuned by a gate voltage applied to the island.

The third part of the total Hamiltonian is given by
\begin{equation}
	H_\text{tun}=\sum_{\eta \vec k\vec k'\sigma}t_{\eta}a_{\eta\vec k\sigma}^\dagger a_{\text{I}\vec k'\sigma}+\text{H.c.}
\end{equation}
It describes tunnel processes of electrons between the island and the reservoirs where $t_\eta$ is the tunnel matrix element which characterizes the coupling strength between the island and the leads. The tunnel matrix element is linked to the tunnel resistance of the junction
\begin{equation}
R_\eta^{-1}=4\pi e^2\rho_\text{N,I}\rho_{\text{N},\eta}|t_{\eta}|^2/\hbar.
\end{equation}

\section{\label{sec:method}Real-time diagrammatic transport theory}
We describe  transport of charge and heat through the superconducting single-electron transistor by means of a real-time diagrammatic approach~\cite{konig_zero-bias_1996,konig_resonant_1996,konig_quantum_1999}. It allows one to describe transport far from equilibrium for a small, interacting quantum system coupled to noninteracting reservoirs. It takes into account interactions of the quantum system exactly while performing a systematic perturbation theory in the system-reservoir coupling. The method has been formulated initially for systems with normal metal reservoirs and has subsequently been extended to account for superconducting reservoirs as well~\cite{governale_real-time_2008,*governale_erratum:_2008,kamp_phase-dependent_2019}. The central idea of the real-time diagrammmatic approach is to integrate out the noninteracting degrees of freedom in the reservoirs to arrive at a generalized master equation for the elements $P^{\chi_1}_{\chi_2}=\bra{\chi_1}\rho_\text{red}\ket{\chi_2}$ of the reduced density matrix of the quantum system $\rho_\text{red}$. In the stationary state, the generalized master equation reads
\begin{equation}
	0=-\frac{i}{\hbar}\left(E_{\chi_1}-E_{\chi_2}\right)P^{\chi_1}_{\chi_2}+\sum_{\chi'_1\chi'_2} W^{\chi_1\chi'_1}_{\chi_2\chi'_2}P^{\chi'_1}_{\chi'_2},
	\label{eq:MEQ}
\end{equation}
where the first term describes the coherent evolution of the quantum system while the second term accounts for dissipative processes due to the coupling to the reservoirs. The generalized transition rates $W^{\chi_1\chi'_1}_{\chi_2\chi'_2}$ are given by irreducible self-energy blocks of the dot propagator on the Keldysh contour. In the following, we are going to consider the transition rates to lowest order in the tunnel coupling. This implies that we account for ordinary tunneling of quasiparticles as well as for Andreev-like processes where electronlike quasiparticles are converted into holelike quasiparticles while creating or annihilating Cooper pairs~\cite{governale_real-time_2008,kamp_phase-dependent_2019}. However, our approximation does not include the coherent transfer of Cooper pairs across the system such that we do not obtain any finite Josephson currents in our calculations.

In the following, we focus on transport for gates voltages close to the point $n_g=N$ where $N$ is an integer. For temperatures small compared to the charging energy $E_C$, transport takes place via sequential transitions from the ground state in which the island is occupied with $N$ electrons to the two excited states in which the island hosts $N\pm1$ electrons. The two excited states are quasi degenerate and, therefore, can form coherent superpositions in the stationary state. Hence, the relevant elements of the reduced density matrix are given by the three occupation probabilities $P_{N-1}$, $P_N$ and $P_{N+1}$ and the two coherences $P^{N+1}_{N-1}$, $P^{N-1}_{N+1}$. In this regime, the generalized master equation can be cast into a physically intuitive form by introducing a pseudospin degree of freedom that describes the coherent superposition of island states with different charge number. To this end, we define the probabilities $\mathcal P_g$ and $\mathcal P_e$ to find the island in its ground and excited state, respectively, as
\begin{equation}
	\mathcal P=\left(\begin{array}{c}\mathcal P_e \\ \mathcal P_g\end{array}\right)=\left(\begin{array}{c} P_{N-1}+P_{N+1}\\ P_N\end{array}\right),
\end{equation}
and the pseudospin $\vec I$ with components
\begin{align}
	I_x&=\re P^{N+1}_{N-1},\\
	I_y&=\im P^{N-1}_{N+1},\\
	I_z&=\frac{P_{N-1}-P_{N+1}}{2}.
\end{align}
The master equation for the occupation probabilities takes the form
\begin{equation}
	0=\sum_{\eta=\text{L,R}} \left[\left(\begin{array}{cc} -\mathcal W^-_\eta & \mathcal W^+_\eta \\ \mathcal W^-_\eta & -\mathcal W^+_\eta \end{array}\right)\mathcal P + 4\mathcal X^-_\eta\left(\begin{array}{c} 1\\-1\end{array}\right)\vec I\cdot \vec n_\eta\right].
	\label{eq:FEQ}
\end{equation}
The first term on the right-hand side describes tunneling events between the island and lead $\eta$ with transition rates $\mathcal W^\pm_\eta$. Their precise form is given in Appendix~\ref{app:rates}. The second term accounts for a coupling of the occupation probabilites to the pseudospin due to Andreev tunneling processes. The associated rate $\mathcal X^-_\eta$ is given in Appendix~\ref{app:rates}. The vector 
\begin{equation}
	\vec n_\eta=\left(\begin{array}{c}\cos(\phi_\eta-\phi_\text{I}) \\ \sin(\phi_\eta-\phi_\text{I}) \\ 0 \end{array}\right)
\end{equation}
accounts for the phases of the superconducting order parameters in the leads and on the island.

The master equation for the pseudospin takes a Bloch-type form and reads
\begin{equation}
	0=\vec A-\frac{\vec I}{\tau}+\vec I\times\vec B.
\end{equation}
The first term, 
\begin{equation}
	\vec A=\sum_{\eta=\text{L,R}} \left(\mathcal X^-_\eta\mathcal P_e+\mathcal X^+_\eta\mathcal P_g\right)\vec n_\eta
	\label{eq:Iacc}
\end{equation}
describes the accumulation of pseudospin via Andreev tunneling events. The second term accounts for the relaxation of the pseudospin due to quasiparticle tunneling with a relaxation time given by
\begin{equation}
	\frac{1}{\tau}=\sum_{\eta=\text{L,R}} \mathcal W^-_\eta.
\end{equation}
Finally, the last term describes a precession of the pseudospin in an effective magnetic exchange field given by
\begin{equation}
	\vec B=\sum_{\eta=\text{L,R}} B_\eta\vec n_\eta+4E_C(N-n_g)\vec e_z.
\end{equation}
The exchange field consists of two different contributions. The first one is oriented in the $x-y$ plane. Its direction is controlled by the superconducting phases while its absolute value is given by the principal value integral
\begin{align}
	B_\eta =-\frac{1}{2\pi e^2R_\eta}&\int^\prime d\omega d\omega' f^+_\eta(\omega)f^+_\text{I}(\omega')\rho_\eta(\omega) \rho_\text{I}(\omega') \nonumber \\
	& \times \frac{\Delta_\eta}{\omega}\frac{\Delta_\text{I}}{\omega'}\frac{1}{E_C+\omega+\omega'},
\end{align}
where $f^+_\eta(\omega)=[\exp(\omega/(\kBT_\eta))+1]^{-1}$ denotes the Fermi function of the reservoirs and the island.
It arises from virtual Andreev tunneling processes between the island and the reservoirs which renormalize the energies of the island states relative to each other. Interestingly, the exchange field affects transport already to lowest order in the tunnel coupling because the associated precession angle arises as the product of the exchange field and the dwell time of electrons on the island.
The second contribution to the exchange field is due to the energy difference between the states with $N+1$ and $N-1$ eleectrons on the island and points along the $z$ axis.

The charge and heat current through the system can be obtained by computing the corresponding current rates. They are given by the generalized transition rates multiplied with the amount of charge or heat that is transferred in the associated tunneling event. We remark that the real-time diagrammatic approach conserves charge and energy currents automatically. In the following, we define the charge current as the current flowing between the left lead and the island. It is given by
\begin{equation}
	I^e=-2e(-B_{\text{L},y}I_x+B_{\text{L},x}I_y+\mathcal W^-_\text{L}I_z).
	\label{eq:Ie}
\end{equation}
Similarly, we define the heat current as the current flowing between the left lead and the island. It reads
\begin{equation}
	I^h=E_C(-\mathcal W^-_\text{L}\mathcal P_e+\mathcal W^+_\text{L}\mathcal P_g+4\mathcal X^-_\text{L}\vec I\cdot \vec n_\text{L}).
		\label{eq:Ih}
\end{equation}
At this point, we emphasize again that we consider sequential tunneling only and disregard higher-order tunneling processes. Therefore, processes which transfer Cooper pairs coherently from the left to the right superconductor are not included in our theoretical description. As a result, the charge and heat currents in Eq.~\eqref{eq:Ie} and Eq.~\eqref{eq:Ih} depend only on the phase difference $\phi_\text{R}-\phi_\text{L}$ across the entire junction. Therefore, we set without loss of generality $\phi_\text{L}=-\phi_\text{R}=\phi/2$ and $\phi_\text{I}=0$.

\section{\label{sec:results}Results}
In this section, we are going to analyze the charge and heat currents flowing through the system in response to an applied temperature and phase bias. We will first focus on the
regime of a temperature cascade $(T_\text{L}>T_\text{I}>T_\text{R})$  while in the second part we turn to a situation where the two superconducting reservoirs are kept at the same temperature such that $(T_\text{L}=T_\text{R} \neq T_\text{I})$.

\subsection{Temperature Cascade}
\begin{figure}
	\includegraphics[width=0.95\columnwidth]{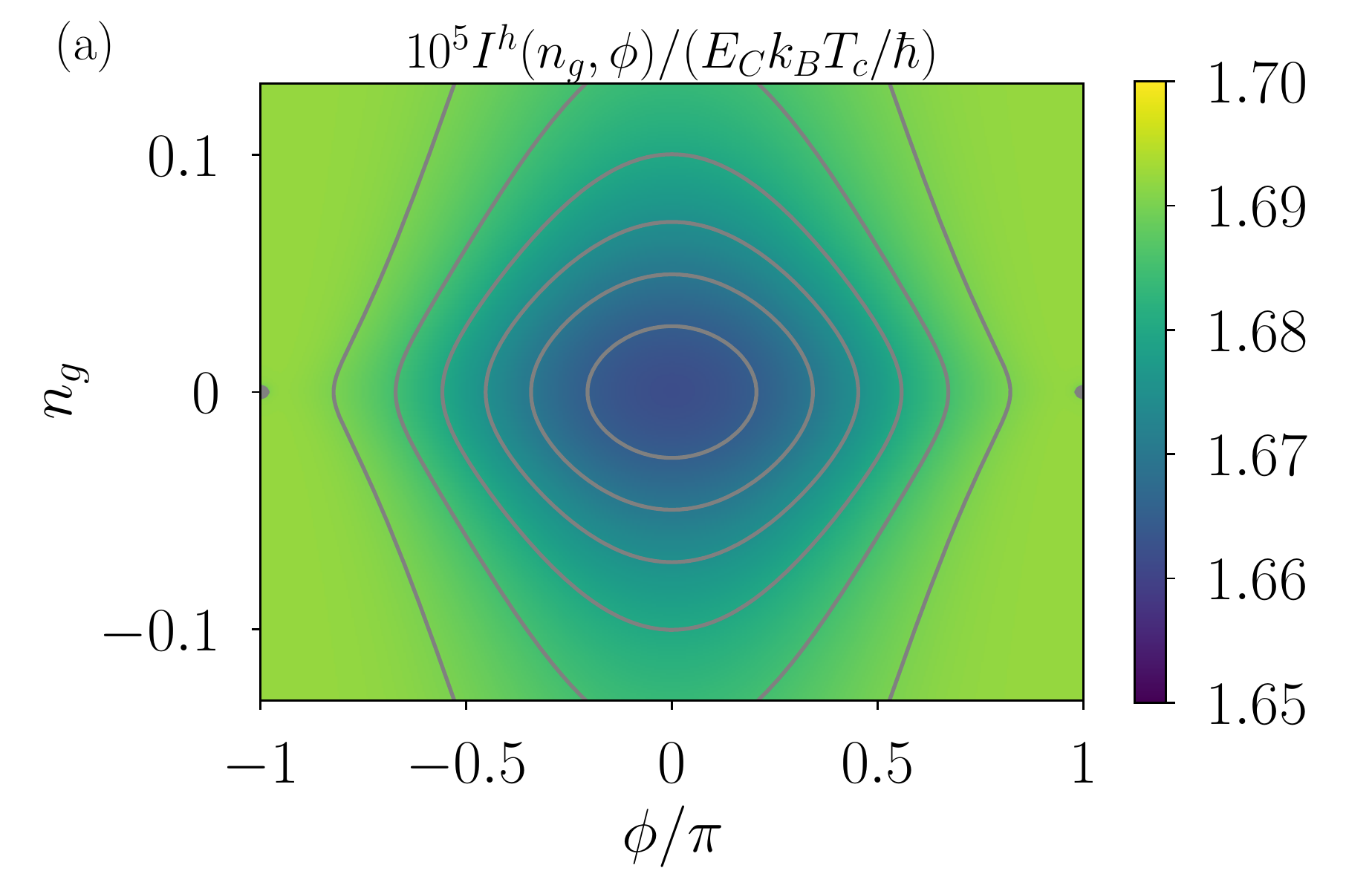}
	\includegraphics[width=0.95\columnwidth]{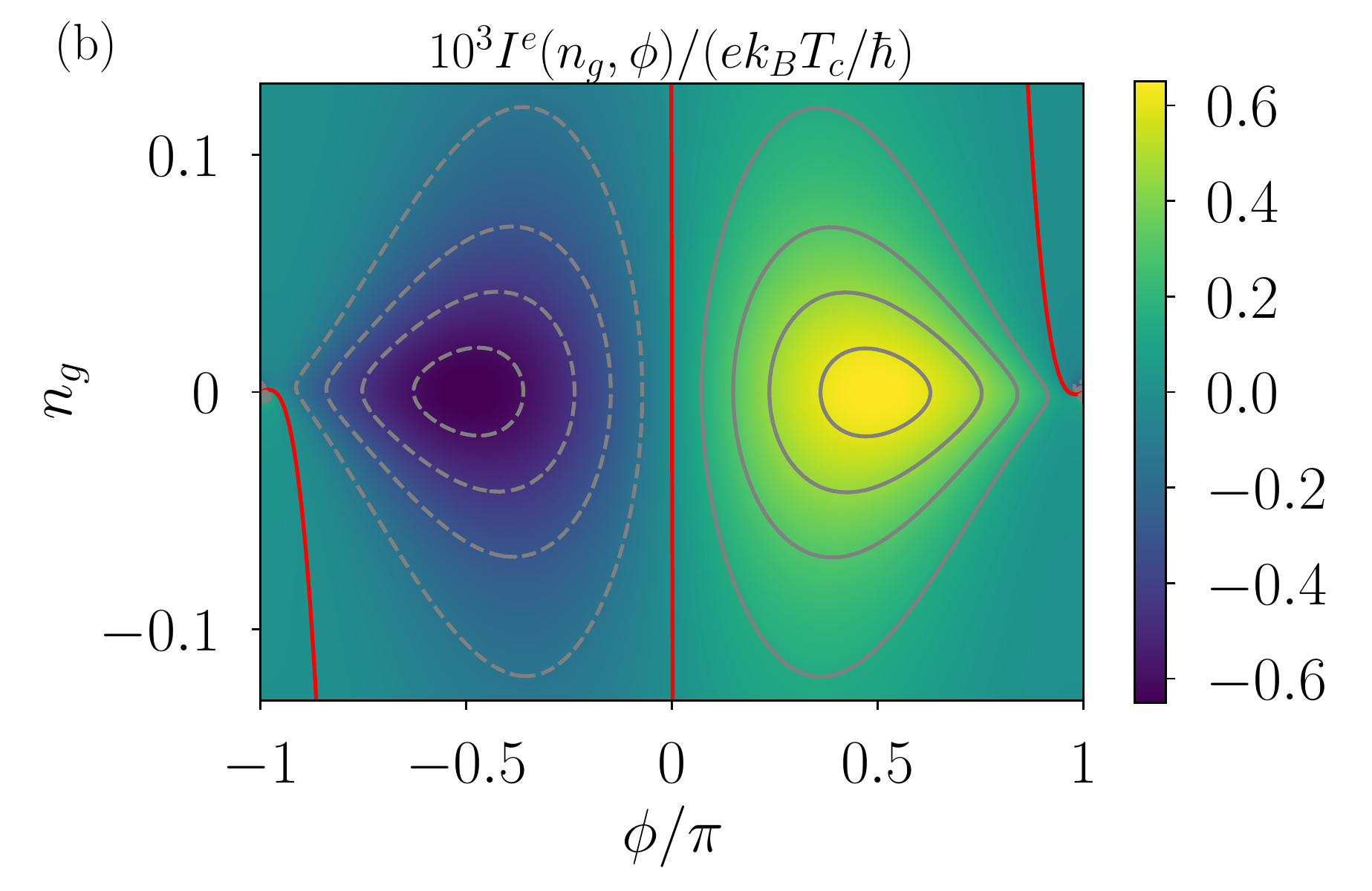}
	\caption{\label{fig:T_ladder}(a) Heat current $I^h$  and (b) charge current $I^e$ as a function of detuning $n_g$ and phase bias $\phi$. Parameters are chosen as $
	E_c=k_B T_c/2$, $T_\text{L}=T_c/4$, $T_\text{I}=T_c/8$, $T_\text{R}=T_c/16$, $R_\text{L}=R_\text{R}=0.1\cdot 2\pi e^2/\hbar$ and $\Gamma_\eta=\Delta_\eta/10$. The red line indicates a vanishing electrical current.}
\end{figure}
We consider the case of a finite temperature bias between the left and the right superconductor. In particular, we assume a temperature cascade with $T_L>T_I>T_R$. This is motivated by the fact that in experiment one would heat one of the two superconducting terminals which in turn would lead to an increased island temperature, e.g., due to leakage heat currents. 

Let us start our discussion by analyzing the heat current that flows in response to the temperature bias. As can be seen in Fig.~\ref{fig:T_ladder}, the heat current is always larger than zero, i.e., flows from hot to cold in agreement with the second law of thermodynamics. The magnitude of the heat current can be manipulated by both the phase bias and the gate voltage. It fulfills the symmetry relation
\begin{equation}
I^h(n_g,\phi)=I^h(-n_g,-\phi),
\end{equation}
because $I_x$ is invariant under a simultaneous sign change of $n_g$ and $\phi$ while $I_y$ and $I_z$ change sign.

The phase dependence of the heat current exhibits a global minimum at $\phi=0$. This behavior is similar to the phase-dependent conductance of a tunnel junction~\cite{maki_entropy_1965,maki_entropy_1966} and differs from the behavior of the heat current in a superconductor-quantum dot hybrid which exhibits a minimum at phase difference $\phi=\pi$. The phase-dependence of the current in our single-electron transistor can be understood from the fact that the island has a continuous density of states such that the system acts like a series of two tunnel junctions rather than the superconductor-quantum dot hybrid where transport is strongly influenced by the discrete level structure of the quantum dot.

Compared to the average value of the heat current, its phase modulation is rather weak. The reason for this is that normal tunneling of quasiparticles is enhanced by the BCS density of states which becomes very large close to the gap edge. The rates for Andreev-type processes which are responsible for the phase-dependent contribution to heat transport instead involve a factor $\rho_\eta(\omega)\Delta_\eta/\omega$ which lacks the enhancement at the gap edge.

When $n_g$ is varied away from $n_g=N$, we find that the phase-dependence of the heat current becomes suppressed. The physical reason for this suppression lies in the fact that the phase dependence arises from Andreev-type processes which induce transitions between the ground state of the island with $N$ electrons and the coherent superposition of the two excited states with $N-1$ and $N+1$ electrons. The generation of these coherent superpositions is most effective when the two states are degenerate in energy, i.e., for $n_g=N$ and gets suppressed by a finite detuning $n_g$. 
As the number of thermally excited quasiparticles is exponentially suppressed by $E_C/\kBT$, the heat current drops exponentially as a function of this quantity.

We now turn to a discussion of the charge current that flows in response to the temperature bias. As shown in Fig.~\ref{fig:T_ladder}(b), the thermoelectric current is in general finite and can be manipulated by both the phase bias $\phi$ and the equilibrium occupation of the island $n_g$. Just as for the heat current, we find that it is suppressed exponentially by the ratio $E_C/\kBT$. The direction of the current can be controlled by the phase bias which is in contrast to the behavior of the heat current discussed above. The reason for this is that the charge current is sensitive to whether transport is dominated by electron- or hole-like quasiparticles. In contrast, the heat current is insensitive to the nature of the quasiparticles that contribute to it. Just as for the heat current, we find that detuning the system away from $n_g=N$ weakens the phase dependence of the charge current. While the heat current is a symmetric function of phase bias and detuning, the heat current is antisymmetric and fulfills the symmetry relation
\begin{equation}
I^e(\phi,n_g)=-I^e(-\phi,-n_g),
\end{equation}
which again follows from the symmetry properties of $\vec I$ and $\vec B$ under a simultaneous sign change of $\phi$ and $n_g$. 
Interestingly, the charge current does not vanish at the particle-hole symmetric point $n_g=0$ but rather at the implicit solution of the equation
\begin{equation}
\frac{W_L^-}{W_R^-}=\frac{B_L \sin{(\alpha-\phi/2)}}{B_R \sin{(\alpha+\phi/2)}}\,,
\end{equation}
where $\alpha=\arctan I_x/I_y$ as follows from setting both, the charge current from the island to the left and to the right superconductor equal to zero.

At this point, let us emphasize that the charge current constitutes a finite thermoelectric current in the vicinity of the particle-hole symmetric point. In particular, we remark that in calculation the transition rates of the generalized master equation, we have assumed a detuning $E_{N+1}-E_{N-1}$ which is of the order of the tunnel coupling and could, therefore, be neglected when evaluating the Fermi functions. Hence, one could naively expect that the thermoelectric current in our system vanishes identically. However, virtual tunneling of quasiparticles between the dot and the reservoirs gives rise to a level renormalization which manifests itself as the exchange field. As the energies of the states with $N-1$ and $N+1$ electrons are renormalized differently with respect to the state with $N$ electrons, particle-hole symmetry is effectively broken and a finite thermoelectric current can arse. This also explains why at phase difference $\phi=0$ and $\phi=\pi$ there is a vanishing thermoelectric current. Here, the exchange field is collinear to the pseudospin and, thus, has to impact on the transport properties of the system. The observed effect is similar to what has been predicted for superconductor-quantum dot hybrids where a similar level renormalization also gives rise to thermoelectric effects in the vicinity of the particle-hole symmetric point~\cite{kamp_phase-dependent_2019}.

\subsection{Inversion-symmetric temperature bias}
\begin{figure}
	\includegraphics[width=0.99\columnwidth]{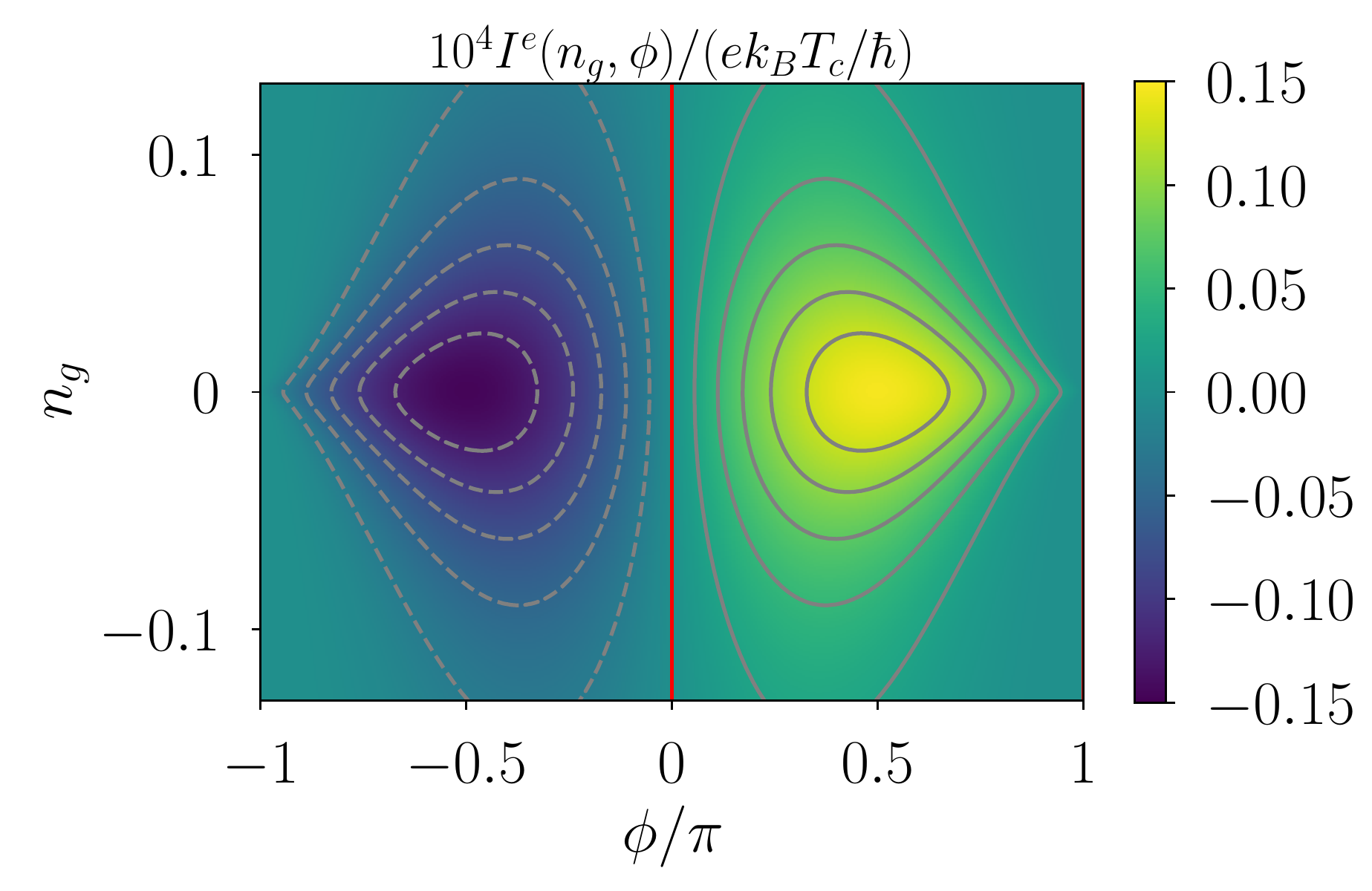}
	\caption{\label{fig:T_sym}Electrical current $I^e$ as a function of detuning $n_g$ and phase bias $\phi$. The temperatures are chosen as $T_\text{L}=T_c/8$, $T_\text{I}=T_c/16$, $T_\text{R}=T_c/8$ while the remaining parameters are chosen as in Fig.~\ref{fig:T_ladder}. The red line indicates a vanishing electrical current.}
\end{figure}

We now consider a situation where both reservoirs are kept at the same temperature while the island is at a different temperature, $T_\text{L}=T_\text{R}\neq T_\text{L}$. In the absence of superconductivity, this would imply that the junction is inversion-symmetric such that both heat and charge currents vanish. Remarkably, we find that in the presence of superconductivity this pictures changes due to the possibility to have a finite phase bias across the junction.

We start our analysis by looking at heat transport. We find that the heat current through the system vanishes independently of the phase bias and detuning. This is a consequence of the symmetry of the system as any heat flow through the system would imply that heat flows from the cold to the hot in one of the two tunnel junctions in violation of the second law of thermodynamics.

We now turn to the electrical current. As can be seen in Fig.~\ref{fig:T_sym}, we find a finite thermoelectric current when a finite phase bias is applied to the junction. The charge current obeys the symmetry relation $I^e(\phi,n_g)=-I^e(-\phi,n_g)$. This implies that there is no charge transport for $\phi=0$ which indicates the importance of a finite phase bias to get a finite charge current. We remark that the finite thermoelectric current is a nonlinear effect that depends quadratically on the applied temperature bias.

\begin{figure}
	\includegraphics[width=0.95\columnwidth]{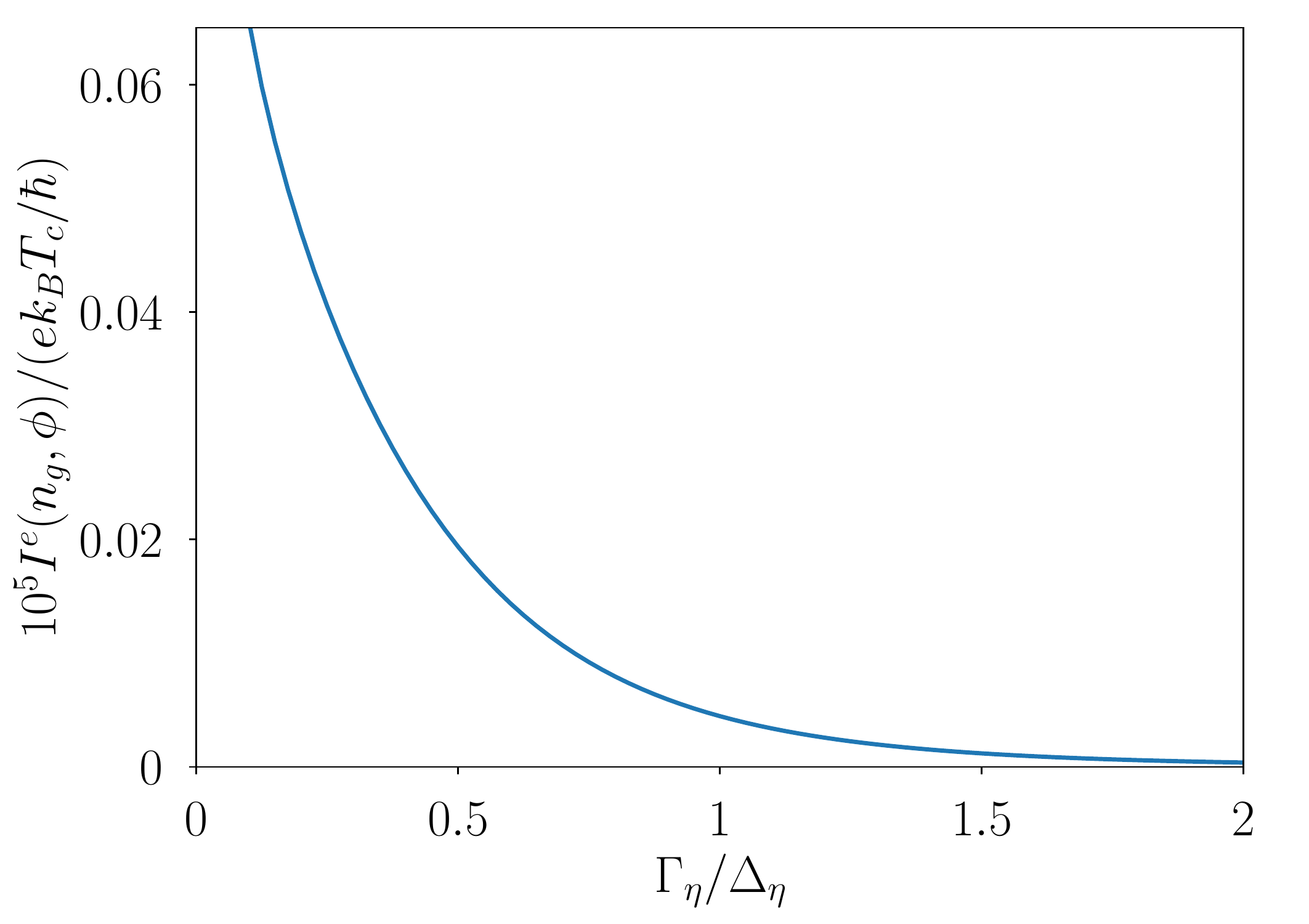}
	\caption{\label{fig:Dynes}Effect of the Dynes parameter $\Gamma_\eta$ on the charge current $I^e$ evaluated at $\phi=\pi/2$ and $n_g=0$ for $E_C=k_B T_c$. The remaining parameters are chosen as in Fig.~\ref{fig:T_sym}. As the Dynes parameter increases, the thermoelectric current gets suppressed.
	}
\end{figure}

The physical picture underlying the thermoelectric transport in the symmetric configuration is as follows. The temperature bias between the island and the two reservoirs drives particles across the two tunnel junctions. Importantly, the phase difference experienced by the quasiparticles is opposite across the two junctions. As a result, Andreev-type processes which convert electron-like quasiparticles to hole-like quasiparticles dominate on one junction while Andreev processes that convert hole-like quasiparticles to electron-like quasiparticles dominate at the other junction. Consequently, this gives rise to a finite electrical current across the system. To confirm this picture, we have calculated the thermoelectric current for different values of the Dynes parameter. Increasing the Dynes parameter effectively leads to a closing of the superconducting gap and, thus, to a suppression of Andreev processes. As can be seen in Fig.~\ref{fig:Dynes}, the charge current is indeed suppressed when the Dynes parameter is increased.

The occurrence of a finite thermoelectric current can also be derived readily from the generalized master equation. Since the two superconductors have the same temperature, $T_\text{L}=T_\text{R}$, the generalized transition rates and the exchange field contributions from the two leads are equal to each other, 
\begin{equation}
\mathcal W^{\pm}_\text{L}=\mathcal W^{\pm}_\text{R},\qquad \mathcal X^{\pm}_\text{L}=\mathcal X^{\pm}_\text{R}, \qquad B_\text{L}=B_\text{R}.
\end{equation}
Therefore, the accumulation term points along the pseudospin accumulation points along the $x$ axis,
\begin{equation}
\left(\frac{d\vec I}{dt}\right)_\text{acc}=2\left(\mathcal X^-_\text{L}\mathcal P_e+\mathcal X^+_\text{L}\mathcal P_g\right)\cos\frac{\phi}{2} \vec e_x, 
\end{equation}
and is collinear with the exchange field which for $n_g=0$ is given by
\begin{equation}
\vec B=2B_\text{L}\cos\frac{\phi}{2} \vec e_x.
\end{equation}
Therefore, the pseudospin acquires only a finite $x$ component while its $y$ and $z$ component vanish. As a result, the charge current is simply given by
\begin{equation}
	I^e =2eB_{\text{L},y} I_x.
\end{equation}
which is in general finite and proportional to $\sin\phi$.

We conclude by commenting on the transport effects for the inversion-symmetrically biased system from a thermodynamic perspective. At first glance, it seems counterintuitive to have a finite thermoelectric current without any accompanying heat flow. We remark, however, that we do not apply any bias voltage against which the thermoelectric current could perform work. Therefore, a finite charge current without corresponding heat current does not violate the laws of thermodynamics. Incorporating a finite voltage in our theoretical description would involve counting the number of Cooper pairs transferred across the system~\cite{governale_real-time_2008}. In this case, we expect nontrivial transport features such as the AC Josephson effect to play a role which is, however, beyond the scope of our present work.

\section{\label{sec:conclusions}Conclusion}
We have investigated thermally-driven transport in a superconducting single-electron transistor by means of a real-time diagrammatic approach that accounts for strong Coulomb interactions and performs a systematic expansion in the tunnel coupling. We analyzed heat and charge currents for the case that a finite temperature bias is applied across the system. We find that both currents depend in a nontrivial fashion on the gate voltage and phase bias between the superconducting reservoirs. Remarkably, the system shows a finite thermoelectric effect in the vicinity of the particle-hole symmetric point. It arises due to level renormalizations caused by virtual tunneling processes in the presence of strong Coulomb interactions.

Furthermore, we analyzed transport in a configuration with an inversion-symmetric temperature bias. While the heat current vanishes in this setting due to second law of thermodynamics, a finite, nonlinear charge current arises when a finite phase bias is applied across the system. It is due to Andreev-like processes which experience opposite phase differences across the two junctions of the single-electron transistor.

\acknowledgments
We acknowledge financial support from the Ministry of Innovation NRW via the ``Programm zur Förderung der Rückkehr des hochqualifizierten Forschungsnachwuchses aus dem Ausland''.

\appendix
\section{\label{app:rates}Transition rates}
In the following, we provide the analytical expressions for the various transition rates that occur in the generalized master equations for the dot occupations and the pseudospin. In particular, for $\eta=\text{L,R}$, they are given by
\begin{widetext}
\begin{align}
	\mathcal W^-_\eta&=\frac{1}{e^2R_\eta}\int d\omega \rho_\eta(\omega-E_C)\rho_\text{I}(\omega)f^+_\eta(\omega-E_C)f^-_\text{I}(\omega),\label{eq:Wm}\\
	\mathcal W^+_\eta&=\frac{1}{e^2R_\eta}\int d\omega \left[\rho_\eta(\omega+E_C)\rho_\text{I}(\omega)f^-_\eta(\omega+E_C)f^+_\text{I}(\omega)+\rho_\eta(\omega-E_C)\rho_\text{I}(\omega)f^-_\eta(\omega-E_C)f^+_\text{I}(\omega)\right], \label{eq:Wp} \\
	\mathcal X^+_\eta&=-\frac{1}{e^2R_\eta}\int d\omega \frac{\Delta_\eta}{E_C-\omega}\frac{\Delta_\text{I}}{\omega}\rho_\eta(E_C-\omega)\rho_\text{I}(\omega)f^+_\eta(E_C-\omega)f^+_\text{I}(\omega), \label{eq:chip} \\
	\mathcal X^-_\eta&=-\frac{1}{2e^2R_\eta}\int d\omega \frac{\Delta_\eta}{E_C+\omega}\frac{\Delta_\text{I}}{\omega}\rho_\eta(\omega+E_C)\rho_\text{I}(\omega)f^-_\eta(\omega+E_C)f^+_\text{I}(\omega), \label{eq:chim}
\end{align}
\end{widetext}
where $f^-_\eta(\omega)=1-f^+_\eta(\omega)$.

% \bibliography{/home/bjoern/LaTeX/Bibtex/Meine_Bibliothek.bib}

\end{document}